\documentclass[mathleft
]{an}
\usepackage{graphicx}
\usepackage{times}
\usepackage{uffa}

\newcommand{\procspie}{Proc.~SPIE}
\newcommand{\nat}{Nature}
\newcommand{\aaps}{A\&AS}%
\newcommand{\aap}{A\&A}
\newcommand{\memsai}{Mem.~Soc.~Astron.~Italiana}
\newcommand{\glog}{log\,g}
\newcommand{\COBOLD}{{\sf CO$^5$BOLD}}
\newcommand{\cobold}{\COBOLD}

\newcommand{\xx}{\ensuremath{\mathrm{1D}_{\mathrm{LHD}}}}
\newcommand{\mD}{\ensuremath{\left\langle\mathrm{3D}\right\rangle}}
\newcommand{\loggf}{\ensuremath{\log\,gf}}
\def\kms{$\mathrm{km\, s^{-1}}$}
\newcommand{\mlp}{\ensuremath{\alpha_{\mathrm{MLT}}}}
\newcommand{\teff}{T$_{\rm eff}$}

\usepackage{natbib}
\bibpunct{(}{)}{;}{a}{}{,}
\begin{document}

\Pagespan{789}{}
\Yearpublication{2011}%
\Yearsubmission{2010}%
\Month{11}%
\Volume{999}%
\Issue{88}%

\title{Extremely metal-poor stars in SDSS fields,\thanks{based on spectra obtained with X-Shooter
at the 8.2m Kueyen ESO telescope, GTO programmes
085.D-0194 and 086.D.0094}}

\author{P. Bonifacio\inst{1}\fnmsep\thanks{Corresponding author:
  \email{Piercarlo.Bonifacio@obspm.fr}\newline}
\and  E. Caffau\inst{2,1}
\and  P. Fran\c cois\inst{1}
\and  L. Sbordone\inst{3,1}
\and  H.-G. Ludwig\inst{2,1}
\and  M. Spite\inst{1}
\and  P. Molaro\inst{4}
\and  F. Spite\inst{1}
\and  R. Cayrel\inst{1}
\and  F. Hammer\inst{1}
\and  V. Hill\inst{5}
\and  M. Nonino\inst{4}
\and  S. Randich\inst{6}
\and  B. Stelzer\inst{7}
\and  S. Zaggia\inst{8}
}
\titlerunning{Extremely metal-poor stars from SDSS}
\authorrunning{P. Bonifacio et al.}
\institute{
GEPI, Observatoire de Paris, CNRS, Universit\'e Paris Diderot, Place
Jules Janssen, 92190
Meudon, France
\and 
Zentrum f\"ur Astronomie der Universit\"at Heidelberg, Landessternwarte, K\"onigstuhl 12, 69117 Heidelberg, Germany
\and 
Max-Planck Institut f\"ur Astrophysik, Karl-Schwarzschild-Str. 1, 85741 Garching, Germany
\and 
Istituto Nazionale di Astrofisica,
Osservatorio Astronomico di Trieste,  Via Tiepolo 11,
34143 Trieste, Italy
\and
Universit\'e de Nice Sophia Antipolis, CNRS,
Observatoire de la C\^ote d'Azur, Laboratoire Cassiop\'e e, B.P. 4229, 06304 Nice Cedex 4, France
\and
Istituto Nazionale di Astrofisica,
Osservatorio Astrofisico di Arcetri, Largo E. Fermi 5, 50125 Firenze, Italy
\and
Istituto Nazionale di Astrofisica,
Osservatorio Astronomico di Palermo, Piazza del Parlamento 1, 90134 Palermo, Italy
\and
Istituto Nazionale di Astrofisica,
Osservatorio Astronomico di Padova Vicolo dell'Osservatorio 5, 35122 Padova, Italy
}

\received{20 Dec 2010}
\accepted{   Jan 2011}
\publonline{later}

\abstract{%
Some insight on 
the first generation of stars can be obtained
from the chemical composition of their direct
descendants, extremely metal-poor stars (EMP),
with metallicity
less than or equal to 1/1000 of the solar metalllicity. 
Such  stars are exceedingly 
rare, the most successful surveys, for this purpose, have so far provided 
only about 100 stars with 1/1\,000 the solar metallicity
and  4 stars with about 1/10\,000 of the
solar metallicity.  
The Sloan Digital Sky Survey has the 
potential to provide a large number of 
candidates of extremely low metallicity. 
X-Shooter has the unique capability
of performing the necessary follow-up 
spectroscopy providing accurate metallicities and abundance ratios 
for several elements (Mg, Al, Ca, Ti, Cr, Sr,... ) 
for EMP candidates. 
We here report on the results for the first two stars observed in the course 
of our franco-italian X-Shooter GTO.
The two stars were targeted to be of metallicity around --3.0,
the analysis of the X-Shooter spectra showed them to be of metallicity
around --2.0, but with a low $\alpha$ to iron ratio, which explains
the underestimate of the metallicity from the SDSS spectra.
The efficiency of X-Shooter allows an {\em in situ} study of the outer Halo,
for the two stars studied here we estimate distances of 3.9 and 9.1 Kpc,
these are likely the most distant dwarf stars studied in detail to date.
}

\maketitle
\sloppy
\urldef\xshome\url{http://www.eso.org/sci/facilities/paranal/instruments/xshooter/index.html}. 

\section{Introduction}

Starting from the very simple chemical
composition of the Universe emerged from the 
Big Bang nucleosythesis (hydrogen, helium and a trace of lithium,
which we shall call the {\em primordial} chemical composition),
the more complex elements have built up as products
of several generations of stars.
Our Galaxy holds the fossil record of this build-up:
the chemical composition of the oldest stars.
The very first generation of stars necessarily 
had the primordial chemical composition.
If in this first generation, stars of  low
mass had been formed, they would still be shining
on the Main Sequence today. 
The search for stars of extremely low metallicity
could allow us to find such  stars, if they ever existed.
If not, we can then conclude that
the first generation stars were all massive and are thus
extinct. The lowest metallicity stars that
can be found will tell us when the star-formation
switched from producing only massive
stars to producing stars of all masses, like observed
today. Finally the chemical composition of the most
metal-poor stars, that are the direct descendants
of the first generation,  provides us indirect
information on the first generation and the chemical
elements that they produced.

The data of the
Sloan Digital Sky Survey \citep[SDSS,][]{york,dr6} 
provides an excellent opportunity for
searching for primordial stars.
Thanks to
its high efficiency, along with intermediate resolution and broad
spectral  coverage, X-Shooter \citep{dodorico}  
offers the unique possibility of follow-up
spectroscopy of the EMP candidates 
in the magnitude range $18\le g\le 20$. 
The surface density of EMP stars in 
this 
magnitude range is so low (less than one
per square degree) that it is not worth to use
multi-object spectrographs.

\section{Target selection}

Our targets were selected among the stars
which have spectra in the Sloan Digital Sky Survey.
The two stars here described were extracted from
the analysis of SDSS Data Release 6 \citep[SDSS-DR6][]{dr6}, and
have colours compatible with the Halo Turn-Off. 
Temperatures are derived from $g-z$ colour by using the colour-temperature
calibration presented in \citet{Ludwig08}.
For the X-Shooter sample we concentrated on the stars
fainter than $g=18$ with colours $-0.3\le g-z \le 0.7$
and $u-g > 0.7$.
The metallicities were estimated using 
a version  of the automatic
abundance  determination code of \citet{BC03},
tailored to the resolution and coverage of the SDSS
spectra.
All the stars with estimated metallicities below --2.0
where then visually inspected.

\begin{table*}
\caption{Coordinates and SDSS photometry of the target stars\label{tabco}}
\begin{tabular}{ccccrrrr}
\hline
Name  & $\alpha(2000)$ & $\delta(2000)$ & $g$ & $(g-z)_0$ & $(i-z)_0$ & $\rm E(B-V) $ & $V_r$ \\ 
      &                &                &     &           &           &               &  \kms \\  
\hline
 SDSS J135046.74+134651.1 &  13:50:46.740 & 13:46:51.00 & 18.29 & 0.320& $0.048$ & 0.027 &  $-331\pm 1$ \\
 SDSS J135516.29+001319.1 &  13:55:16.290 & 00:13:19.00 & 18.97 & 0.131&$-0.004$ & 0.040 &  $-9 \pm 1$\\
\hline
\end{tabular}
\end{table*}

\begin{figure}
\includegraphics[width=80mm,clip=true]{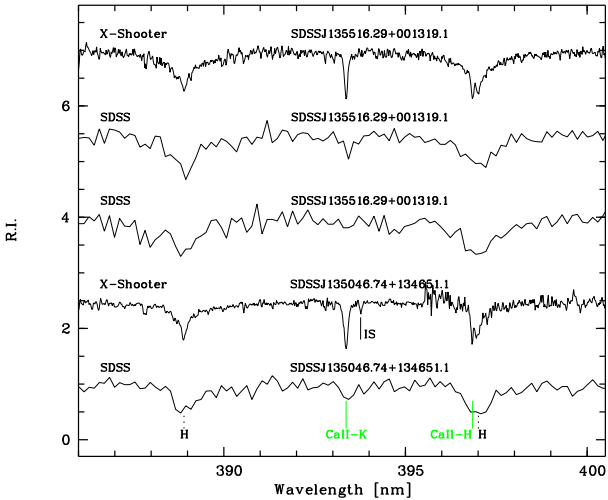}
\caption{
The SDSS 
spectra of the program stars in the region
around the \ion{Ca}{ii} K line compared to the X-Shooter
spectra.
The fact that this line is weaker than
the neighbouring H lines is a diagnostic
for extremely low metallicity. 
In the X-Shooter  spectrum of SDSS J135046.74+134651.1
we labelled ``IS'' the weak interstellar \ion{Ca}{ii} K line.
}
\label{sdssspectra}
\end{figure}

The two stars selected are
SDSS J135046.74+134651.1 and SDSS J135516.29+001319.1,
whose basic data are summarized in Table \ref{tabco}.
From the analysis of the SDSS spectra SDSS J135046.74+134651.1 
was expected to have a metallicity
[M/H]$\sim    -3$\footnote{[X/H] = log(X/H)-log(X/H)$_\odot$} .
As a first try we did not select particularly extreme objects,
the goal being to demonstrate the possibility of using
X-Shooter spectra for abundance analysis.
For SDSS J135516.29+001319.1 there are two SDSS spectra,
one provided a metallicity around --3, the other around --2.
In Fig. \ref{sdssspectra} the SDSS spectra in the region 
around the Ca{\sc ii} K line   are  shown and compared
to the X-Shooter spectra, to emphasize the increase in 
resolution and S/N ratio.  

\section{Observations and data reduction}

Both stars were observed in visitor mode with the IFU in stare
mode\footnote{For a full description of X-Shooter observing modes and related
acronyms see the instrument documentation at 
\xshome}. 
In our case the use of the IFU was not aimed
at obtaining any spatial information (the stars are point sources),
but to collect as many photons as possible with resolution
corresponding to that of a slit of 0\farcs{6}, i.e. R=7900\relax\ in the
UVB arm and R=12600 \relax in the VIS arm.
SDSS J135046.74+134651.1 was observed on April 7th 2010, with
a single integration of 5400 s, in the IR arm the integration time
was divided into 6 DITs of 900\,s each.
 SDSS J135516.29+001319.1 was observed on May 17th 2010, with two integrations
of 3150\,s each, in the IR arm each integration was split into
4 DITs of 787\,s.
The different integrations were co-added.
The spectra were reduced using the X-Shooter pipeline \citep{goldoni}.
The IR arm did not contain any useful information, the present paper
is based on the analysis of the UVB and VIS arms.
The UVB arm covers the range 300-559.5\,nm and the VIS arm
559.5-1024\,nm. 

The signal to noise ratios achieved were 
S/N=37   at 415\,nm and 38 at 660\,nm for  SDSS J135046.74+134651.1
and S/N=18  at 415\,nm and 20 at 660\,nm for  SDSS J135516.29+001319.1.

\subsection{Analysis}

Our abundance analysis is based on 1D ATLAS model atmospheres. We began
with atmospheres 
computed with version 9 of the ATLAS code \citep{kurucz93,kurucz05}
on Linux \citep{sbordone04,sbordone05}. 
In this version of ATLAS the line opacity is treated
through Opacity Distribution Functions (ODFs) and we
used the ones computed by \citet{2003IAUS..210P.A20C},
for a microturbulent velocity of 1\,\kms.
For SDSS J135516.29+001319.1,
given that its abundance pattern is significantly different
from that assumed in the computation of the ODFs, we also computed
a specific model atmosphere with version 12 of the
ATLAS code, which uses opacity sampling \citep{kurucz05,Castelli05},
and our final analysis is based on this model.
Convection was treated in the mixing-length approximation
with a mixing-length parameter \mlp = 1.25.

We also used a 3D hydrodynamical model atmosphere
(3D model here after) computed with the \cobold\ code, (for
details see \citealt{freytag10}) to compute  3D corrections. This
\cobold\ model has  parameters \teff / \glog /[M/H]=6320K/4.5/--2.0, 
and the series of 20 selected snapshots
covers 2.5\,h. As a reference 1D models we used two
1D model atmospheres, 
\mD\ and \xx :
\mD\ is obtained from the 3D model by averaging all snapshots horizontally
at surfaces of equal Rosseland optical depth;
\xx\ is computed with the code LHD, that employs
the same microphysics and radiative transfer scheme as
the \cobold\ code.
For both these models, in the spectral
synthesis we used a microturbulence of 1.5\kms, and
the mixing-length parameter of the LHD model was fixed at 0.5. 

The 3D corrections are defined as in \citet{caffau10},
the difference of the abundance derived from the 3D model
and the one derived from the 1D reference models:
$\Delta_{\rm gran}={\rm A(X)_{\rm 3D}-A(X)_{\rm \mD}}$
to isolate the granulation effects, due to horizontal fluctuations; 
$\Delta_{\rm LHD}={\rm A(X)_{\rm 3D}-A(X)_{\rm \xx}}$
to measure the effects that both the horizontal fluctuation and the temperature
structure have on the abundance determination.

In the spectrum synthesis we used the atomic parameters used
in the ``First Stars'' survey 
\citep{paperV,paperVIII,paperXII}. 
The reference solar iron abundance 7.52 is 
from \citet{caffau10}. For the other elements
the solar reference is from \citet{lodders09}.

We measured the equivalent widths (EW) of the
lines in the observed spectra using the IRAF task {\tt splot}
and derived the abundance with the code WIDTH \citep{kurucz05,castelliw}.
We used the line positions of the lines used for the abundance
analysis in the UVB spectrum to derive a radial velocity of the stars.
The results are given in Table \ref{tabco} along with
 the error  
on the mean of the velocities.
We did not observe any radial velocity standards, thus
we do not have a good estimate of the external error on these radial
velocities. 
For star SDSS J135046.74+134651.1 the SDSS spectrum
provides a radial velocity of $(-299\pm 24)$\kms, compatible
with our measurement at less than 2 $\sigma$.
For star SDSS J135516.29+001319.1, there are two SDSS spectra,
which provide $(-43 \pm 34)$\kms and $(-33 \pm 30)$\kms,
although both measurements are compatible with ours, it is
possible that the star is a radial velocity variable. 
These comparisons
suggest that the external error is not larger than a few \kms .

The large negative radial velocity of SDSS J135046.74+134651.1
allows to clearly resolve the interstellar \ion{Ca}{ii} K line,
as shown in Fig. \ref{sdssspectra}. The line has an EW of 1.6\,pm,
such a weak line is expected, given the low reddening of this star.
For star SDSS J135516.29+001319.1 the low radial velocity 
implies that the photospheric line is contaminated by the interstellar
absorption, however, since this star has an even lower
reddening, we expect the interstellar line to be even weaker, 
thus the contamination should be less than 1\%.


\section{SDSS J135046.74+134651.1}

\subsection{Stellar parameters}

From the $(g-z)_0$
colour we derived an effective temperature of 6284\,K.  
The  colour excess provided by the SDSS catalogue for this star
is given in Table \ref{tabco} and is
is based on the reddening maps of \citet{schlegel}. It is very low
and ignoring it would result in an effective temperature
lower by 65\,K.
As a check, we also derived T$_{\rm eff}$ by fitting the H$\alpha$ wings. 
The theoretical profiles were
computed using the self broadening theory of \citet{barklem} and
the Stark broadening of \citet{stehle}.
As done in \citet{sbordone10}, we used  the CIFIST grid 
of \cobold\ models \citet{Ludwig09}, a
grid of \xx\ models with a mixing
length parameter of 0.5, a gravity of 4.5, and metallicity
-2.0 and --3.0,  and an analogous 
grid of ATLAS 9 models.
From the \cobold\ models 
we obtain a temperature of 6420\,K (see Fig.\,\ref{star1_ha}), from
the \xx\ models 6240\,K, and from the ATLAS models
6400\,K.  
The difference between \xx\ and ATLAS models is not
surprising, given the different temperature structures of the models.
The photometric and H$\alpha$-based 
temperatures are consistent within $\sim 100$\,K, and each of them 
has an associated error which is again of about 100\,K. Moreover, 
the accuracy of the present analysis is limited mainly
by the limited S/N ratio and resolution of the spectra, rather 
than by the temperature uncertainty. We thus decided
to adopt the photometric temperature.
The gravity has been fixed at \glog =  4.5, based on 
the \ion{Fe}{i} and \ion{Fe}{ii} equilibrium.
The resolution of X-Shooter, even with the IFU, and
the achieved S/N ratio,  do not allow to measure
weak lines, on the linear part of the curve of growth.
Therefore the microturbulence cannot
be derived from the observation.  
This parameter is  not a really independent 
parameter in 1D analysis, and it can be calibrated
as a function of effective temperature and, mainly,
surface gravity. 
We decided to apply the relation
of \citet{edvardsson93} which yields a
microturbulence of 1.5\,\kms.
It must be noted that in this way microtrubulence
and surface gravity are strictly correlated.
Any change in surface gravity implies
a change in microturbulence, which implies
a change in the abundances of both 
Fe{\sc i} and Fe{\sc ii}, and thus a change in gravity. 
Therefore the finally adopted values  had to be determined iteratively.

\subsection{Abundances}

The derived abundances
are given in Table\,\ref{tab_star1abbo}, while in 
Table\,\ref{tab_star1lines} all the lines considered are listed with the 
atomic parameters, the equivalent width (EW), the abundance derived from the photometric temperature
and from the H$\alpha$ 3D best fit, the 3D corrections, and the abundance
derived after applying the 3D corrections.

The star is metal-poor, [Fe/H]$=-2.3$, but
not around $-3.0$, as expected from the
analysis of the SDSS spectrum. The striking thing
is that the iron content is similar to the
abundance of $\alpha$-elements. This is unusual because
the metal-poor Halo stars display generally  an enhanced
abundance of $\alpha$-elements.
This expected overabundance of $\alpha$-elements
is in fact folded into our method
of analysis of SDSS spectra where metallicity is derived also 
from  lines of $\alpha$-elements.
In fact at low metallicity the dominant feature is the \ion{Ca}{ii} K line.
It is thus not surprising that such a non-$\alpha$
enhanced star is estimated to have a too low iron 
abundance
from the SDSS spectra.

The elements detectable in the observed spectrum are Fe (both \ion{Fe}{i} and \ion{Fe}{ii}),
Mg, Si, and Ca (both neutral and singly ionised).

\begin{table}
\caption{Chemical abundances from the 1D ATLAS abundance analysis of SDSS J135046.74+134651.1.}
\label{tab_star1abbo}
\begin{tabular}{lllrc}
\hline
El. & [X/H]& [X/H]  & N$^\dag$ & ${\rm A_\odot^\ddag}$\\
    & ${\rm T_{\rm eff}}=6284$\,K &${\rm T_{\rm eff}}=6420$\,K \\ 
\hline
\ion{Mg}{i}  & $-2.30\pm 0.06$ & $-2.15\pm 0.06$ &  2 & 7.54\\
\ion{Si}{i}  & $-2.54$         & $-2.43$         &  1 & 7.52\\
\ion{Ca}{i}  & $-2.15$         & $-1.99$         &  1 & 6.33\\
\ion{Ca}{ii} & $-2.21\pm 0.21$ & $-2.12\pm 0.19$ &  4 & 6.33\\
\ion{Fe}{i}  & $-2.33\pm 0.20$ & $-2.17\pm 0.20$ & 12 & 7.52\\
\ion{Fe}{ii} & $-2.28$         & $-2.25$         &  1 & 7.52\\
\hline
\end{tabular}

$^\dag$ {Number of lines}\\
$^\ddag$ {A(X) = log(X/H) + 12}
\end{table}

We  apply the 3D corrections, $\Delta_{\rm LHD}$, to the 1D analysis 
based on the temperature derived from the photometry.
$\Delta_{\rm LHD}$ corrections are negative, meaning that the abundance derived
from 3D model atmosphere is smaller than the one derived from 1D reference model.
This is due to the fact that hydrodynamical models are usually cooler than hydrostatical models
in the outer part of the atmosphere.
The lines formed in the  outermost layers are thus more
affected by the cooling effect. Usually lines of neutral elements
with low excitation energy, show the largest effects.
$\Delta_{\rm gran}$ is close to zero, meaning that for the observed lines
the horizontal fluctuations do not play a determinant role, 
the strongest effect is of almost 0.2\,dex for
the line of \ion{Fe}{ii}.


\subsection{Li abundance}

\begin{figure}
\includegraphics[width=80mm,clip=true]{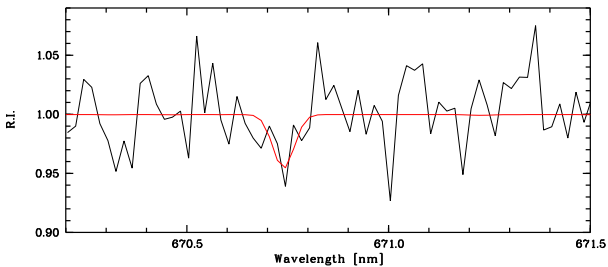}
\caption{
The fit of the Li doublet at 670.7\,nm with a grid of ATLAS+SYNTHE profiles. 
As evident from the feature on the blue
side of the Li, the Li line is comparable with the noise of the spectrum.
}
\label{star1_li}
\end{figure}

The Li feature at 670.7\,nm is not detected, 
its EW being at the level of the noise, so that we consider the
abundance derived as an upper limit. 
With a 1D grid of
ATLAS+SYNTHE synthetic profiles we fitted the observed
profile (see  Fig.\ref{star1_li}). We  determined the EW of the best fit and
used this EW to derive the 3D-NLTE abundance of Li by
using the fitting function of \citet{sbordone10}, which  yields A(Li)=2.38. This value
is compatible with the level of the
Spite plateau \citep{spite82,sbordone10}.

\section{SDSS J135516.29+001319.1}

\subsection{Stellar parameters}

The effective temperature 
from the $(g-z)_0$ colour is extremely high, 
\teff=6765\,K.   
Also for this star the reddening (see Table \ref{tabco})
is low, ignoring the reddening correction would provide
a temperature which is only 106\,K lower, therefore still extremely high.
According to published stellar isochrones,
such a hot star should be  very young, unless it is a Blue Straggler star, 
or lies on the Horizontal Branch.
We note however that the evolutionary
tracks of \citet{piau}, computed with
CNO enhanced chemical composition, predict
such a hot Turn-Off for a population of
12\,Gyrs. 
The colours of this star could in fact suggest a CNO-enhanced
metal-poor star. 

The 3D fit of the H-$\alpha$ profile gives 
a considerably lower effective temperature of 6300\,K.
Among the other colours, $(g-r)_0$ and $(r-i)_0$
support the high \teff\ indicated by $(g-z)_0$;
the $(i-z)_0$ colour is considerably
sensitive  to the surface gravity, it
supports the high temperature
for \glog = 4.5, but for \glog=4.0 it
implies \teff=6300\,K.

We therefore decided to adopt an effective temperature
of 6300\,K, which is consistent with the 
Halo Turn-Off. 
The iron ionisation equilibrium implies \glog = 3.9 and
the microturbulence derived from the \citet{edvardsson93} calibration
is 2.3 \kms .
We note that the inconsistency between
colour and H$\alpha$, as well as the possible
inconsistency of $(g-z)_0$ and $(i-z)_0$, is a hint of a peculiar spectral energy distribution.

\subsection{Abundances}

The abundances  derived assuming both effective temperatures
(photometric and H$\alpha$) are provided in Table
\ref{tab_star2abbo}.
Among the $\alpha$ elements Si and Mg appear underabundant
with respect to iron, but it should be noted that the measurement
is based on a single line for each of the two elements.
Given this unusual chemical composition, different
from the ones available in the ODFs computed by
\citet{2003IAUS..210P.A20C}, and the importance of
Mg as electron donor, we computed models 
with version 12 of the ATLAS code, using the desired chemical
composition. The abundances were then recomputed with this
new model. The differences with respect to those
obtained from an ATLAS model computed with version
9 and non-$\alpha$-enhanced ODFs are very small.
At these low metallicities and high gravities, even consistent variations
of Mg abundance have minor effects on the model
structure.

The \ion{Ca}{ii}  abundance in Table \ref{tab_star2abbo}
is based on two lines of the IR triplet. It must be remarked
that the \ion{Ca}{ii} K line is not compatible with
this abundance and is consistent with a Ca abundance
which is lower by 0.7\,dex.
Since the \ion{Ca}{ii} K line is the main abundance
indicator at low metallicity for the SDSS spectra,
this explains why we had estimated a much lower
metallicity than what results from the analysis
of the X-Shooter spectrum.
Also the \ion{Ca}{i} line, implies an
abundance which is 0.5\,dex {\em higher}
than the \ion{Ca}{ii} IR triplet. 
It should be noted that according to the NLTE computations
of \citet{Mashonkina} the corrections for
\ion{Ca}{ii} triplet lines are always negative,
those for \ion{Ca}{i} 422\.6\, nm are always positive.
Thus the cause of this discrepancy should not be 
due to NLTE effects, which would rather tend to increase
the discrepancy.

We could not convincingly detect the \ion{Li}{i} 670.7 nm feature.
The EW upper limit is 5.4\,pm
corresponding to A(Li)$<$2.98 at a temperature of 6765\,K
and 2.63 at 6300\,K.

We inspected carefully the spectrum for any sign of carbon
enhancement, looking at the G-band and all the \ion{C}{i}
lines which are observed in carbon-enhanced metal-poor
stars (see e.g. \citealt{behara}), but could not detect any.

\begin{table}
\caption{Chemical abundances from the 1D ATLAS abundance analysis for SDSS J135516.29+001319.1.}
\label{tab_star2abbo}
\begin{tabular}{lllrc}\hline
El.      & [X/H] & [X/H] & N & ${\rm A\odot}$\\
         &  ${\rm T_{\rm eff}=6765}$\,K & ${\rm T_{\rm eff}=6300}$\,K &  &  \\
\hline
\ion{Mg}{i}  & $-1.85        $ &  $-2.08        $   &    1 &  7.54\\
\ion{Al}{i}  & $-1.34        $ &  $-1.66        $   &    1 &  6.47\\
\ion{Si}{i}  & $-1.95        $ &  $-2.19        $   &    1 &  7.52\\
\ion{Ca}{ii} & $-1.31\pm 0.25$ &  $-1.59\pm 0.03$   &    2 &  6.33\\
\ion{Ti}{ii} & $-0.84\pm 0.53$ &  $-1.33\pm 0.49$   &    7 &  4.90\\
\ion{Cr}{i}  & $-0.82        $ &  $-1.39        $   &    1 &  5.64\\
\ion{Fe}{i}  & $-1.29\pm 0.31$ &  $-1.74\pm 0.39$   &   12 &  7.52\\
\ion{Fe}{ii} & $-1.40\pm 0.27$ &  $-1.78\pm 0.30$   &    3 &  7.52\\
\ion{Sr}{ii} & $-1.11\pm 0.20$ &  $-1.65        $   &    1 &  2.92\\
\ion{Ba}{ii} & $-1.52        $ &  $-2.10        $   &    1 &  2.17\\
\hline
\end{tabular}
\end{table}

\section{Discussion}

These first observations of metal-poor 
SDSS candidates with X-Shooter
allows us to estimate the strengths and
limitations of this instrument for this application.
The great strength is the high efficiency of X-Shooter,
which has allowed us to obtain a workable spectrum 
of a star of 19th magnitude in less than two hours.
The use of the IFU \citep{IFU} as image slicer
has been successful. The transmission of the IFU
in the UVB arm is lower than at longer wavelengths.
The opportunity of using the IFU  
rather than a slit of 0\farcs{6} depends of course on the
seeing. In the case of the observations described here
the seeing was always between 1\farcs{0} and 1\farcs{2},
the advantage of using the IFU is unquestionable.

The main limitation of X-Shooter for abundance analysis
is the resolution, which is too low to allow us to 
measure lines on the linear part of the curve of growth.
Lines on the flat part of the curve of growth are little
sensitive to abundance, so that a large error is always
associated to abundances derived from such lines.
The situation is better for lines which are even stronger
and lie in the damping part of the curve of growth, although
in this case the damping parameters should be accurately known.
For metal-poor dwarf stars, however, 
even the \ion{Ca}{ii} K line  has weak 
wings and is usually on the flat part
of the curve of growth. A partial compensation is provided by the very
large spectral coverage provided by X-Shooter, which allows
to measure many features, improving the abundance accuracy.
This, however, for metal poor stars, applies only to iron, 
since the other species are represented by only one or few
lines.

The two stars studied in these first GTO runs 
did not turn out to be as metal-poor as expected
from the analysis of the SDSS spectra, the main
reason being, in both cases, the weakness of the
\ion{Ca}{ii} K line with respect to the iron lines.
In fact our analysis of SDSS spectra estimates
the metallicity by using all available features
{\em assuming} that $\alpha$ elements are enhanced
by 0.4\, dex over iron.

In the case of SDSS J135046.74+134651.1, 
for the few
measurable elements,  
the abundances appear to be solar-scaled.
The existence of metal-poor stars with low,
almost solar $\alpha$ to iron
ratios was pointed out by
\citet{NS97}, however, the most metal-poor stars of that sample
had a metallicity about --1.0.
Remarkable  known metal-poor $\alpha$-poor stars are  BD+~80$^\circ$~245,
with a metallicity of --1.8 and a logarithmic $\alpha$ to iron
ratio of --0.3 \citep{carney} and CS 22873-139 \citep{spite00} with
a metallicity of --3.4, [Mg/Fe]=--0.04, [Ca/Fe]=+0.16 and [Ti/Fe]=+0.55.
The latter star is a double spectrum spectroscopic binary.
At lower metallicities stars with bizarre
compositions are known, such as HE 1424-0241 \citep{cohen},
with a metallicity around --4.0 and Si underabundant
by 1\,dex, but magnesium ``normally'' enhanced, 
or SDSS J234723.64+010833.4 \citep{lai} which 
has a small underabundance of 
 Mg ([Mg/Fe]=--0.1) but a strong overabundance of  
Ca ([Ca/Fe]=+1.1), at a metallicity of --3.2.

SDSS J135046.74+134651.1 appears to have
an abundance pattern similar to BD+~80$^\circ$ 245,
and CS 22873-139.
The most straight forward interpretation of these $\alpha$--poor 
metal--poor stars is that they were formed in  low-mass
galaxies, satellites of the Milky Way, characterized by a low
or bursting star-formation rate. The original galaxies have been
disrupted due to tidal interaction, and their debris populates
the Halo. It is interesting to point out that the surviving
satellite galaxies have indeed low $\alpha$ to iron
ratios, although only Carina displays essentially
solar $\alpha$ to iron ratios at a metallicity
around --2.0 (see the review of \citealt{THT}).
However, at even lower metallicity 
([Fe/H]=$\sim$3 and below), solar-scaled $\alpha$ element abundances are
detected in stars of Sculptor and Sextans dSph \citep{tafelmeyer10}.
Thus, in a naive interpretation, the majority
of surviving satellites had a higher star formation
rate than Carina, Sculptor, Sextans or the disrupted satellites,
such as the one in which SDSS J135046.74+134651.1
originated.
A more extensive survey searching for EMP
stars in the halo may allow us to quantify the
percentage of these low-$\alpha$ stars.

The case of SDSS J135516.29+001319.1 is less clear.
The discrepancy, in terms of \teff ,  
between several colours and H-$\alpha$
strongly suggests that the spectral energy distribution
is anomalous. It would be tempting to suggest the presence
of a companion emitting a considerable flux in the UV. 
The most likely candidate is  a hot white dwarf,
however it should be about four magnitudes fainter
than the Turn-Off star, thus its effect would be negligible.
The situation could be different if the white dwarf
were surrounded by an accretion disc providing a significant
UV continuum. We are entering here in the domain of
the wild speculation, and not enough observational
data are available to enable us to present a consistent
hypothesis.
The weakness of the \ion{Ca}{ii} K line with respect to
the \ion{Ca}{ii} IR triplet could in fact be explained
by the presence of excess UV flux, however this could
not explain the unusual strength of the \ion{Ca}{i}
resonance line, which is only 29\,nm to the red
of \ion{Ca}{ii} K.
It is also possible that we are observing a binary system
composed of two stars, of different \teff ~ one could immagine
the cooler companion to contribute the \ion{Ca}{ii} IR triplet
and the \ion{Ca}{i} lines, while the warmer one contributes
the \ion{Ca}{ii} K line. In this case however the measurements
suggest a \teff ~ difference of at least 1000\,K, thus the 
luminosities of the two companions have to be carefully 
chosen, in order to provide the observed spectrum.
The quality of the available data is not sufficient
to explore such composite spectrum solutions.
We conclude that this star has a peculiar spectrum, 
possibly composite and its abundances ought
to be regarded as highly uncertain.
Further radial velocities measurements could allow
to decide if the star is a binary.

We want to stress that the unique efficiency of X-Shooter
opens up the possibility to measure the chemical abundances
of the outer Halo {\em in situ}. To estimate the distances
we used the SDSS photometry and the Padova isochrone
of 13 Gyr and metallicity --2.0 \citep{marigo}.
The distances (averaged over all five SDSS bands)
are 3.9 Kpc for SDSS J135046.74+134651.1 and
9.1 Kpc for SDSS J135516.29+001319.1, they are likely the
most distant dwarf stars for which the detailed chemical
composition could be measured.
We used the Besan\c con Galactic model \citep{robin} to
simulate a field of stars with colours and magnitudes
similar to our own.
In the field around SDSS J135046.74+134651.1, the
distribution of radial velocities has mean of 0\kms
with a dispersion of 80~\kms , confirming that 
the star, with radial velocity -331~\kms, has little
to do with the Halo, and strengthening the hypothesis
of its extra-Galactic origin. Instead in the case of
SDSS J135516.29+001319.1 its nearly zero radial velocity
is perfectly consistent with the mean radial velocity
of Halo stars in this field. 


\appendix

\section{Appendix A}

\begin{table*}
\caption{Abundance from each line, from the 1D ATLAS+WIDTH, and 3D corrections for SDSS J135046.74+134651.1.}
\label{tab_star1lines}
\begin{tabular}{lclcrrrrccc}\hline

Element      &$\lambda$ & $\xi$ & \loggf  & EW$^\dag$      & $\Delta_{\rm gran}$ & $\Delta_{\rm LHD}$
 & [X/H]  & [X/H]  & [X/H]\\
             &          & eV    &         & [pm]   &  & &  ${\rm T_{\rm eff}}$\,K & ${\rm T_{\rm eff}}$\,K 
&$_{\rm 1D}+\Delta_{\rm gran}$\\
             &          &       &         &        &  & &     =6284                   &   =6420 \\  
\hline
\ion{Fe}{i}  & 381.5840 & 1.485 &   0.237 &  16.1 & --0.03 & --0.21 & --1.96 & --1.78 & --2.17 \\
\ion{Fe}{i}  & 382.0425 & 0.860 &   0.119 &  15.0 & --0.09 & --0.33 & --2.47 & --2.28 & --2.80\\
\ion{Fe}{i}  & 385.6371 & 0.052 & --1.286 &  10.6 & --0.14 & --0.56 & --2.30 & --2.11 & --2.86 \\
\ion{Fe}{i}  & 386.5523 & 1.011 & --0.982 &   8.6 &   0.00 & --0.34 & --2.17 & --2.01 & --2.51 \\
\ion{Fe}{i}  & 404.5812 & 1.485 &   0.280 &  11.8 &   0.03 & --0.26 & --2.48 & --2.31 & --2.75 \\
\ion{Fe}{i}  & 406.3594 & 1.557 &   0.062 &   9.9 &   0.04 & --0.28 & --2.51 & --2.35 & --2.79 \\
\ion{Fe}{i}  & 413.2058 & 1.608 & --0.675 &   7.0 & --0.05 & --0.32 & --2.40 & --2.26 & --2.72 \\
\ion{Fe}{i}  & 414.3868 & 1.557 & --0.511 &   8.2 & --0.01 & --0.33 & --2.36 & --2.22 & --2.69 \\
\ion{Fe}{i}  & 426.0474 & 2.399 &   0.109 &   6.8 &   0.00 & --0.18 & --2.40 & --2.28 & --2.59 \\
\ion{Fe}{i}  & 438.3545 & 1.485 &   0.200 &  10.6 &   0.03 & --0.33 & --2.65 & --2.49 & --2.98 \\
\ion{Fe}{i}  & 440.4750 & 1.557 & --0.142 &  10.5 &   0.03 & --0.34 & --2.26 & --2.10 & --2.61 \\
\ion{Fe}{i}  & 495.7596 & 2.808 &   0.233 &   8.5 &   0.04 & --0.21 & --2.03 & --1.90 & --2.22 \\
\ion{Fe}{ii} & 516.9033 & 2.891 & --0.870 &   7.7 &   0.18 &   0.16 & --2.28 & --2.25 & --2.12 \\
\ion{Mg}{i}  & 517.2684 & 2.712 & --0.380 &  15.0 & --0.02 & --0.21 & --2.36 & --2.21 & --2.57 \\
\ion{Mg}{i}  & 518.3604 & 2.712 & --0.158 &  20.0 &   0.00 & --0.21 & --2.24 & --2.08 & --2.46 \\
\ion{Si}{i}  & 390.5523 & 1.909 & --1.090 &   8.9 &   0.06 & --0.16 & --2.54 & --2.43 & --2.70 \\
\ion{Ca}{i}  & 422.6728 & 0.000 &   0.240 &  14.1 & --0.10 & --0.42 & --2.15 & --1.98 & --2.57\\
\ion{Ca}{ii} & 393.3663 & 0.000 &   0.134 & 159.5 & --0.05 & --0.19 & --2.46 & --2.31 & --2.65\\
\ion{Ca}{ii} & 849.8023 & 1.692 & --1.312 &  22.1 &   0.09 & --0.25 & --1.92 & --1.85 & --2.17\\
\ion{Ca}{ii} & 854.2091 & 1.700 & --0.362 &  36.2 &   0.01 & --0.20 & --2.35 & --2.27 & --2.55\\
\ion{Ca}{ii} & 866.2141 & 1.692 & --0.623 &  35.3 &   0.02 & --0.21 & --2.11 & --2.05 & --2.32\\
\hline
\end{tabular}

$^\dag$ The relative errors on EWs are of the order of 25\%.
\end{table*}

\begin{figure}
\includegraphics[width=80mm,clip=true]{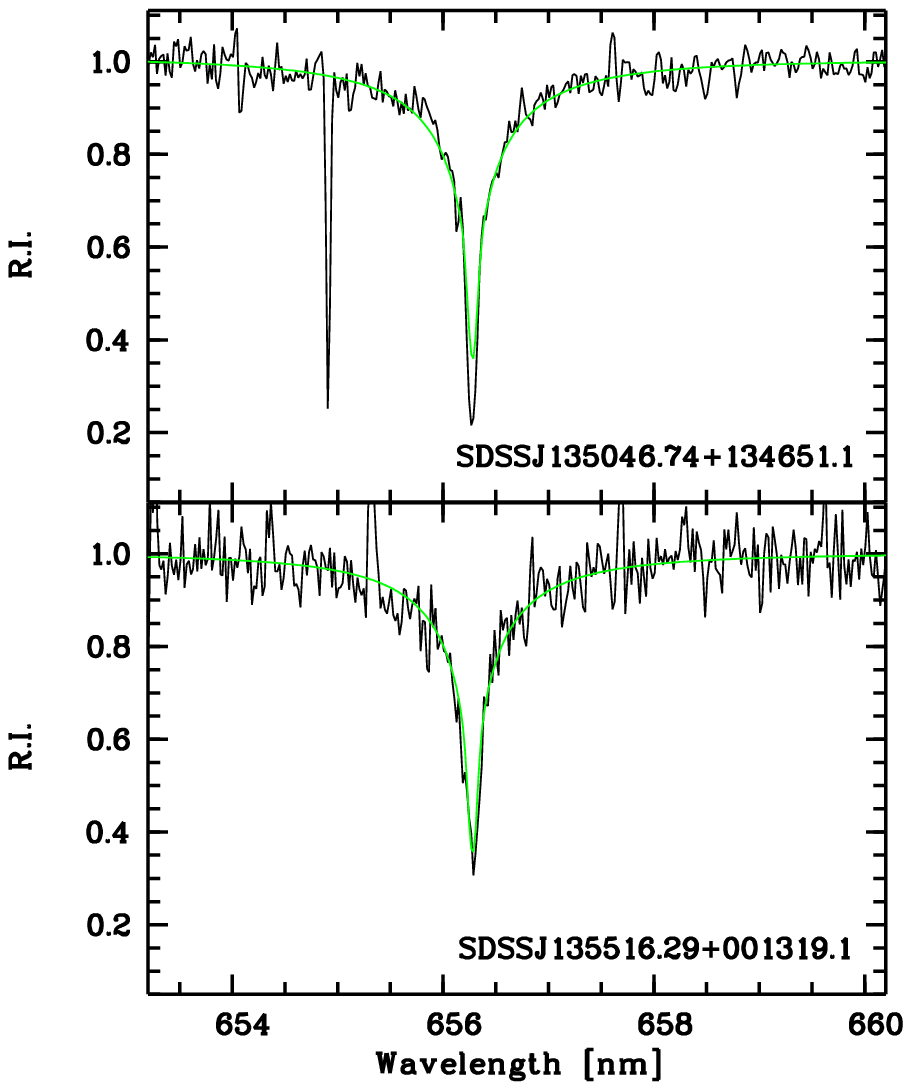}
\caption{
The X-Shooter 
spectrum of SDSS J135046.74+134651.1 (solid black) with 
the best fitting \cobold\ profile (\teff = 6420\,K, solid green) in the top
panel and the same for SDSS J135516.29+001319.1 (\teff = 6288\.K, solid
green) in the bottom panel.
}
\label{star1_ha}
\end{figure}

\end{document}